\documentclass[10pt,conference,a4paper]{IEEEtran}
\usepackage{epsfig}
\usepackage{cite}
\usepackage{amsfonts}
\usepackage{graphics}
\usepackage{amsmath}
\usepackage{amssymb}
\usepackage{latexsym}
\usepackage[]{fontenc}
\usepackage{psfrag}
\usepackage{dsfont}
\usepackage{booktabs}
\usepackage{bm}
\usepackage{algorithm}
\IEEEoverridecommandlockouts
\begin{document}
\title{Accurate Measurement of Power Consumption Overhead During FPGA Dynamic Partial Reconfiguration}

\author{\IEEEauthorblockN{Amor Nafkha, and Yves Louet}

\IEEEauthorblockA{CentraleSup\'elec/IETR, Campus de Rennes \\ 
Avenue de la Boulaie, CS 47601, 35576 Cesson S\'evign\'e Cedex, France.\\
Email: \{amor.nafkha,yves.louet\}@centralesupelec.fr}
}

\maketitle

\begin{abstract}
In the context of embedded systems design, two important challenges are still under investigation. First, improve real-time data processing, reconfigurability, scalability, and  self-adjusting capabilities of hardware components. Second, reduce power consumption through low-power design techniques as clock gating, logic gating, and dynamic partial reconfiguration (DPR) capabilities. Today, several application, \emph{e.g.}, cryptography, Software-defined radio or aerospace missions exploit the benefits of DPR of programmable logic devices. The DPR allows well defined reconfigurable FPGA region to be modified during runtime. However, it introduces an overhead in term of power consumption and time during the reconfiguration phase. In this paper, we present an investigation of power consumption overhead of the DPR process using a high-speed digital oscilloscope and the shunt resistor method. Results in terms of reconfiguration time and power consumption overhead for Virtex 5 FPGAs are shown. 

\end{abstract}


%
\IEEEpeerreviewmaketitle

\section{Introduction}\label{Sec:Intro}
Software Defined Radio (SDR) refers to reconfigurable or re-programmable radios that can perform different functionality with the same hardware \cite{ref0,ref01}. The main objective is to produce communication devices which can support any wireless standards and services. The reconfigurability poses a set of new challenges in terms of real-time processing, power consumption, flexibility, performance. Consequently, many heterogeneous platforms composed of digital signal processors, application specific integrated circuits, and FPGAs have been developed to achieve sufficient processing power and flexibility. To cope with flexibility and computational issues, the concept of FPGA Dynamic Partial Reconfiguration (DPR) can be used \cite{ref1,ref2,ref3,ref4,ref5,ref6}.

The DPR is achieved by loading the partial bitstream of a new design into the SRAM-based FPGA configuration memory and overwriting the current one. Thus the reconfigurable portion will change its behavior according to the newly loaded configuration. In this procedure, the reconfiguration time, $i.e.$ reconfiguration throughput, represents an important and critical design parameter which determines the switching time switching time between the two configuration. This factor must be taken into account in many cases where performance critical applications require fast switching of IP cores. To achieve runtime DPR, authors in \cite{ref2} provide a high-speed interface that transfers the DPR bitstreams into the Internal Configuration Port (ICAP). They measured a maximum DPR speed of 400 MB/s that can be achieved by using Direct Memory Access (DMA) for data transfers between external memory and the ICAP interface (clock speed of 100 MHz). In \cite{overclocking}, authors use an over-clocking ICAP interface (clock speed of 133 MHz) and provide a maximum DPR measured speed of 418.5 MB/s.

To the best of our knowledge, there is a few work which address the experimental measurement and/or the power consumption estimation during the DPR process. In \cite{Bonamy1}, authors presents a detailed investigation of power consumption of the DPR process. However, they use underclocked ICAP interface in order to catch power consumption during DPR given the fact that the partial configuration time is very small. Recently, Texas Instruments provides the Fusion Digital Power Designer software package \cite{TI} with the USB Interface
Adapter EVM \cite{usbevm} to monitor real-time power consumption values. However, this method did not provide high temporal resolution to measure the FPGA power consumption during DPR process. In the present paper, we will present a experimental approach using high-speed digital oscilloscope and the shunt resistor method to provide power consumption during DPR process. One of the main results of this work is that the power consumption overhead, during the FPGA dynamic partial reconfiguration, is low and doesn't exceed 160 mW.    

The paper is structured as follows. Section II provides introduction to different actors providing the utilization of Dynamic Partial reconfiguration concept. In order to make experimental results more representative, variable digital filters are introduced and implemented in Section III. Hardware architecture is provided in section IV. The experimental setup is introduced in Section V. The results of the experiments are presented and discussed in Section VI. Finally, Section VII provides a conclusion.

\section{Partial reconfiguration}\label{sec:PR}
Xilinx company proposes two main FPGA family: Spartan and Virtex devices. Both families can support partial reconfiguration through ICAP component inside the FPGA. The section gives different actors providing the utilization of dynamic partial reconfiguration design approach 

\subsection{The ICAP primitive}
The ICAP allows to access to configuration data and it has the same signaling interface as SelectMAP and can be configured in 8 or 32 bits mode, depending on the target device use. The ICAP primitive is the hardwired FPGA logic by which the bitstream can be dynamically loaded into the configuration memory. For partial reconfiguration management, we use an embedded microprocessor (either MicroBlaze or PowerPC) to transmit the partial reconfiguration bitstream from storage devices to ICAP in order to accomplish the reconfiguration process. The DPR feature allows to partially reconfigure the FPGA internal configuration logic at the runtime. As a consequence, a specific design flow has to be used so as to define the static areas (which will not change at runtime) and the dynamic areas (which can be changed at runtime). This flow allows to create partial bitstreams for the dynamic and static part. These bitstreams are merged to give the global bitstream which is mandatory for the initial global programming of the FPGA. Then the dynamic part can be reconfigured using the partial bitstreams sent through the ICAP component at runtime.

\subsection{Partial reconfiguration manager }
To manage partial reconfiguration, we use an embedded processor soft core, called MicroBlaze, which is a reduced instruction set computer processor optimized for implementation on any Xilinx FPGA. Moreover, Xilinx provides through Embedded Development Kit software an environment tool to connect peripherals to the MicroBlaze and develop application program to drive it. Several communication ports (OPB bus, FSL links, PLB, LMB) are available. As mentioned before, partial reconfiguration is realized through ICAP component inside the FPGA. This reconfiguration is possible by sending partial bitstream to the ICAP. As a consequence, the different bitstreams required have to be stored in external memory like SRAM or internal FPGA memory like BRAM in the order to reduce writing and reading transfer during communication mechanism to ICAP. In this paper, we use the Xilinx ML550 board to make power and time overhead measurements of the dynamic partial reconfiguration process. We have made the choice to store the bitstreams in an internal dual-RAM memory. 

\subsection{Partial reconfiguration modes} 
FPGA dynamic partial reconfiguration can be performed externally, through the JTAG or SelectMap port, or internally through ICAP component. These two modes takes the same configuration files to reconfigure the desired area inside FPGA. A restriction has to be made for the internal reconfiguration, where only partial reconfiguration can be made. Indeed, the management of the process of reconfiguration has to be monitored by a specific interface. For the external configuration, this management could be realized by different components like DSP or GPP. For the internal configuration, this could be managed by a hardware or a software processor as a MicroBlaze or a specific computing unit. As a consequence, for the internal case a global reconfiguration is impossible, after an initialization phase, due to the lost of the control of reconfiguration process it self.

\subsection{Enhancement ICAP architecture}
Considering the inefficiency when the processor moves data to ICAP, we propose a design approach based on an appropriate architecture in charge of the partial bitstreams management through the ICAP. We use the MicroBlaze to catch the order of reconfiguration. Partial bitstreams are stored in a RAM and are accessible by the MicroBlaze through the OPBEMC controller. Then bitstream dataflow is sent to the ICAP component using one of the ICAP32 DMA IP. The ICAP32 DMA is an IP that we have developed in our team in order to take advantage of new 32 bit mode of Virtex 5. The goal is to get closer to the theoretical throughput of the ICAP capacity, 400MByte/s (ICAP running at a maximum frequency of 100MHz). The Microblaze is only used for the setup phase, after that, the IP works alone and acts like a DMA (Direct Memory Access) for bitstreams transfers between BRAM and ICAP component. This IP is connected to the MicroBlaze via the GPIO, and the ICAP component is configured in 32 bits mode. Specific registers store the start addresses and the offset of the physical memory where partial bitstreams are stored in RAM. The MicroBlaze sends the order of reconfiguration to the IP for the selected configuration, and the IP makes the memory transfer to the ICAP itself using a SRAM controller included in ICAP32 DMA component. So data writing phases are realized by words of 32 bits at each FPGA clock cycle.

\section{Variable Digital Filters} \label{sec:VDF}
This section provides a detailed description of the implemented design in order to make different measurements 

\subsection{APT-VDF filter}

\begin{figure}
  \includegraphics[width=1\columnwidth]{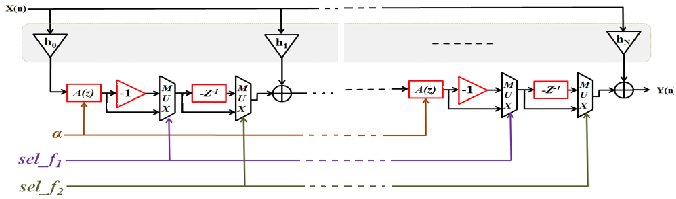}~   
    \caption{APT-based VDF filter architecture}
    \label{fig_apt_vdf}
\end{figure}

All-pass transformation based variable digital filter was initially proposed in \cite{ref5}. Consider a prototype FIR filter with impulse response $h(n)$ and its z-transform
$H(z)$. The APT-VDF version of $H(z)$ is obtained by replacing each unit delay $z$ with the first order all-pass structure, $A(z)$, defined as 
\begin{equation}
A(z) = \frac{-\alpha+z^{-1}}{1-\alpha z^{-1}}, \; |\alpha| <1 
\label{eq1}
\end{equation}
where $\alpha$ is the first order warping coefficient and its value determines the cut-off frequency of the APT-VDF filter. This filter can provide variable frequency response with unabridged control over cut-off frequency. The advantage of the APT-VDF filter is that it allows fine control over cut-off frequency without updating the filter coefficients or structure. The hardware architecture of the APT-VDF filter is depicted in figure \ref{fig_apt_vdf}. 

In Fig. \ref{fig_apt_vdf}, the signals $sel\_f_1$ and $sel\_f_2$ are used to select which filter type will be created:
\begin{itemize}
	\item Lowpass filter when $sel\_f_1 = 0$ and $sel\_f_2=0$,
	\item Bandpass filter when $sel\_f_1 = 1$ and $sel\_f_2=0$,
	\item Highpass filter when $sel\_f_1 = 0$ and $sel\_f_2=1$,
	\item Bandstop filter when $sel\_f_1 = 1$ and $sel\_f_2=1$
\end{itemize}

Consider a $N^{th}$ order lowpass prototype filter with cut-off frequency $f_{co}$, if $0<\alpha<1$, then the resultant cut-off frequencies are lower than $f_{co}$. For $\alpha=0$, the transfer function of the APT-VDF filter is equal to the original transfer function of the prototype filter. If $-1<\alpha<0$, then the resultant cut-off frequencies of the APT-CDF filter are higher than $f_{co}$. The parameter $\alpha$ can be expressed as
\begin{equation}
\alpha =  \frac{sin[(f_{co}-f_{c})\pi/2 ]}{sin[(f_{co}+f_{c})\pi/2 ]}
\label{eq:1}
\end{equation}
where $f_{c}$ is the cut-off frequency of the desired APT-VDF filter. Given a desired frequency response specifications, Matlab filter design tool can be used to obtain the prototype filter coefficients. 

\subsection{Floating to Fixed point conversion}
The APT-VDF architecture has been implemented in a Virtex5 XC5VLX50T FPGA using fixed-point representation. Herein, we consider the generalized fixed-point number representation $[w_l,f_l]$, where $w_l$ and $f_l$ correspond to the word length and the fractional length of the number, respectively. The difference $i_l=w_l-f_l$ is referred to as the integer length of the number. In this paper, the minimum integer word length is calculated under large amounts of simulation data, and the fractional word length are chosen through extensive simulation and tolerable error limit. The root mean square error (RMSE) of output data between floating-point filter and fixed-point filter is evaluated and it is considered as an optimization criteria. Due to the limitation of space, this study will not be presented in this paper. The input, output, and internal signals were represented as an integer and their word length was varied in order to observe the effects of finite precision. Simulations using a serial configuration of the APT-VDF filter at different precisions were made. The results are shown that 12-bit word length and 10-bit fractional precision has significant root mean square error ($RMSE\approx -32 dB$), whereas 16-bit word length and 14-bit fractional precision or higher precision results in very low error ($RMSE\approx -44 dB$).   

\subsection{VHDL Implementation}
The APT-VDF filter has to be implemented in such a way that the design has a static region (SR) and a reconfigurable region (RR) which depends on parameters $sel\_f_1$ and $sel\_f_2$. In this work, we assume that the parameter $\alpha$ is known and fixed. The static region of the APT-VDF filter contains the prototype filter structure and the all-pass filter given the fact that the parameter $\alpha$ is known. Both filters coefficients are given as constants in the hardware architecture. The reconfigurable region of the APT-VDF filter contains the minimum logic that is necessary to each combination of $sel\_f_1$ and $sel\_f_2$ parameters values. The VHDL code of the APT-VDF filter is written in such a way that it is reusable. By changing the filter's order $N$, word length $w_l$, and fractional length $f_l$, an APT-VDF filter of any order and word length can be generated. The ModelSim tool is used to simulate and validate the designed APT-VDF filter. 

\section{Hardware Architecture}

\begin{figure}
  \includegraphics[width=0.8\columnwidth]{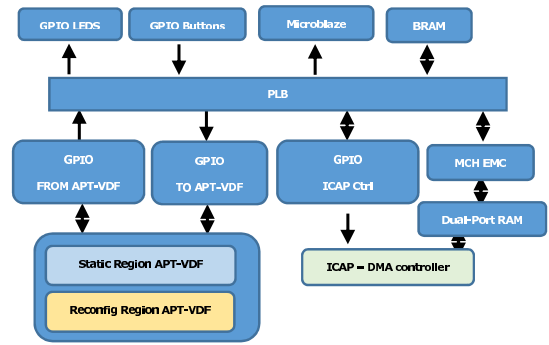}~   
    \caption{Hardware architecture of the designed system}
    \label{archi_apt_vdf}
\end{figure}

The hardware architecture of the implemented APT-VDF filter is shown in Fig. \ref{archi_apt_vdf}. This architecture was built on Xilinx ML550 development board and designed using Xilinx ISE 14.7, XPS 14.7, and PlanAhead 14.7 environments. The architecture contains the following components: Microblaze processor, processor local bus (PLB), memory BRAM, local memory bus (LMB), universal asynchronous receiver transmitter (UART RS232), multi-channel external memory controller (MCH EMC), and five general purpose input/output (GPIOs), true dual-port RAM, internal configuration access port controller, the SR and RR regions of the APT-VDF filter. The static region communicates with Microblaze via GPIO. The UART is integrated to allow the communication between the Microblaze and the RS232 interface of the board. All the peripherals communicate with the Microblaze via the PLB bus. The memory space of the BRAM is configured to be 16 KBytes. The on-chip dual-port RAM is configured to 128 Kbytes and it is used to store the partial bitstream files. The host PC is responsible for transferring those bitstream files to the Microblaze through the UART. The ICAP32 DMA IP is used to take advantage of the 32-bit mode of virtex5. The goal is to get closer to the theoretical throughput of the ICAP capacity, 400MByte/s (ICAP running at a maximum frequency of 100MHz). The Microblaze is only used for the setup phase, after that, the ICAP32 DMA IP controller operates autonomously, and it acts as direct memory access for partial bitstream files transfer between the dual-port RAM and the ICAP component. The ICAP32 DMA IP is connected to the MicroBlaze through a general purpose input/output interface. Two specific registers are used to store the start addresses and the offset of the physical memory where partial bitstream files are stored.

\section{Experimental Setup}
\begin{figure}[t]
    \centering
    \includegraphics[width=0.8\columnwidth]{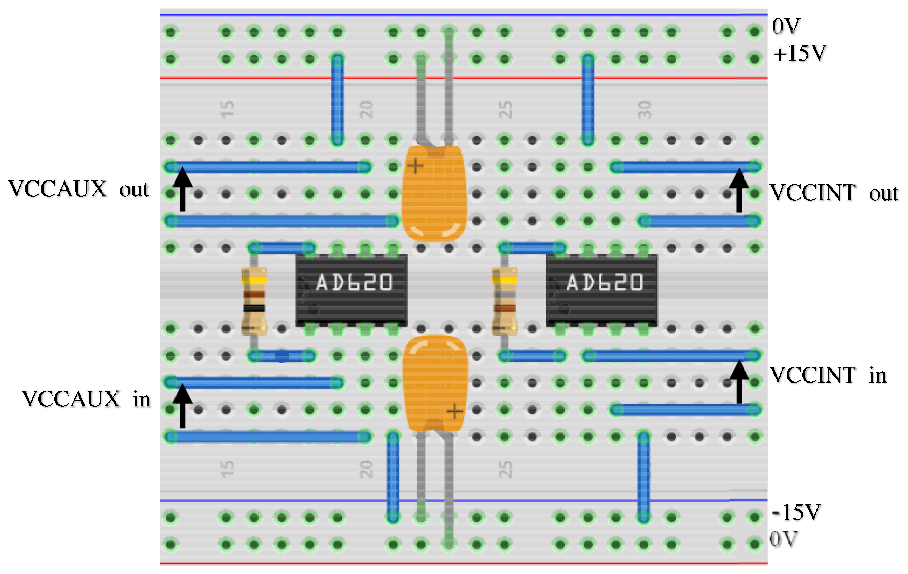}~   
    \caption{AD620 amplification circuit.}
    \label{fig:Sent_Energy}
\end{figure}

The ML550 evaluation boards have been used in this study. It is well suited to power consumption measurement for several reasons. This board provides us with 5 power rails (core, IOs, peripherals) and current sense resistors which would simplify the experimental measurement. The recommended Virtex5 core voltage, designated VCCINT, is 1.0 $\pm$ 10\%V. Depending on the input/output standard being implemented, the Virtex5 I/O voltage supply, designated VCCO, can vary from 1.2 V to 3.3 V. Moreover, Xilinx defines an auxiliary voltage, VCCAUX, which is recommended to operate at 2.5 $\pm$ 10\%V to supply FPGA clock resources. The board hosts a two header connector which provides test points for the ML550 power regulators. Moreover, to measure different currents drained by the FPGA, the ML550 board contains a series of shunt 10 $m\Omega$ $\pm$ 1\% 3W Kelvin current sense resistors on each voltage regulator lines. Thus, the current will be the voltage across the shunt resistor divided by the resistance value itself. Since the sensitivity of the VCCINT, VCCO, and VCCAUX sensors are pretty low (0.5-2mV), voltage amplifiers are needed. To this end, we have used integrated instrumentation amplifiers AD620, from Analog Devices. The AD620 has a feature to increase gain between 1-10.000 times with an external resistor. Those amplifiers allow to regulate the gain only by changing a single resistance, called RG. We wanted to reach a voltage of about 100-400 mV, at the output of the voltage amplifier then, according to the component's availability of the laboratory, was established gains of about 100 and 1000 to amplify the VCCINT and the VCCO voltages using resistances of 483 $\Omega$, and 49 $\Omega$. Fig. \ref{fig:Sent_Energy} shows the developed circuit board to amplify VCCINT and VCCO voltages. The input voltages are connected to a 2 x 13 0.1-inch male header connector which provides test points for the ML550 power regulators. The Tektronix TDS2024C oscilloscope is being used to display the voltages output from the two AD620 amplifiers.
  
\section{Experimental results}

In this section, the implementation results and power consumption measurements of the APT-VDF filter are given with the help of a suitable design example. Herein, all mentioned frequencies are normalized with respect to half of the sampling frequency. Consider a prototype lowpass filter where the cut-off frequency, the transition bandwidth, the passband, and the stopband ripple specifications are 0.08, 0.14, 0.8 dB, and -40 dB, respectively. Using the filter design and analysis tool and direct-form FIR structure, the order of generated prototype filter is equal to $N=21$.

\subsection{DPR area}
\begin{figure}[t]
    \centering
    \includegraphics[width=0.8\columnwidth]{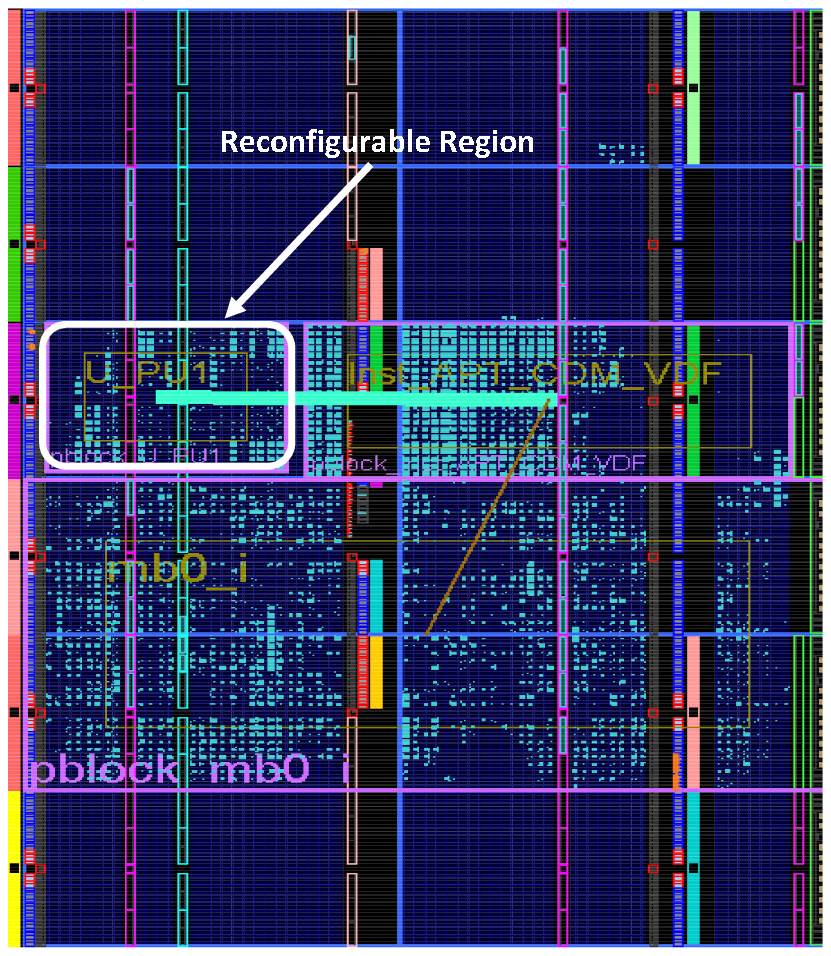}~   
    \caption{Placement of top-level modules.}
    \label{Ahead_fig}
\end{figure}

The partial reconfiguration module has been implemented using the Xilinx PlanAhead tool. The Design view shows the placement of each instance and the post-place information, as
shown in fig. \ref{Ahead_fig}.

Table \ref{tab1} summarizes the key (post-place-and-route) implementation results of the reconfigurable region when the couple of parameters $( sel\_f_1,sel\_f_2)$ is equal to $(0,1)$ and $(1,1)$, respectively. The results are coherent with our expectation given the fact that when $sel\_f_1=0$ and $sel\_f_2=1$, the generated APT-VDF filter is equivalent to highpass filter. However, when $sel\_f_1=1$ and $sel\_f_2=1$, the generated APT-VDF filter is equivalent to a bandstop filter, which requires to insert $N w_l$ delay units.
\begin{table}[!t]
\renewcommand{\arraystretch}{1.3}
\caption{FPGA physical resources of the reconfigurable region}
\label{tab1}
\centering
\begin{tabular}{|c|c|c|}
\hline
\bfseries Resources & \bfseries Highpass  & \bfseries Bandpass\\
\hline
LUT/FD & 315/0 & 336/336\\
\hline
SliceL/SliceM & 67/23 & 77/26\\
\hline
Frames & 16 & 16 \\
\hline
Frame Region & 2 & 2\\
\hline
Bitstream & 94464 Bytes & 94464 Bytes\\
\hline
\end{tabular}
\end{table}

\subsection{DPR timing overhead}
The FPGA devices are composed of columns and rows. The biggest advantage brought by the recent Virtex family is that the reconfiguration can be realized by frames, contrary to Virtex II which imposed to reconfigure the whole columns. This improvement gives more flexibility on the modularity of the design flow during the placement phase of the dynamic and static part on the floorplan of the device. Frame by frame reconfiguration can significantly reduce the size of the generated partial reconfiguration files, and hence the configuration time overhead.

Table \ref{tab2} presents the experimental FPGA reconfiguration time needed to perform APT-VDF filter reconfiguration. Three reconfiguration approaches are used:
\begin{itemize}
	\item Full FPGA reconfiguration: we use JTAG connector to configure the FPGA with a total pre-made bitstream.
	\item External partial FPGA reconfiguration: we use JTAG connector to configure just the reconfigurable region of the APT-VDF filter with a pre-made partial bitstream
	\item Internal partial FPGA reconfiguration: we use the ICAP32 DMA IP to configure just the reconfigurable region of the APT-VDF filter with a pre-made partial bitstream. 
\end{itemize}

\begin{table}[!h]
\renewcommand{\arraystretch}{1.3}
\caption{Experimental results on reconfiguration time overhead}
\label{tab2}
\centering
\begin{tabular}{|c|c|c|}
\hline
\bfseries Reconf. type  & \bfseries Bitstream & \bfseries Reconf. time\\
\hline
Full (JTAG) & 1716 KB  & 3.80 s\\
\hline
Partial (JTAG) & 94.464KB  & 208.9 ms\\
\hline
Partial (ICAP32 DMA) & 94.464KB  & 324 $\mu$s \\
\hline
\end{tabular}
\end{table}

MicroBlaze is only used to configure specific registers which give the beginning address and the offset in RAM where partial bitstreams are stored. As a consequence, we achieve the maximum theoretical input throughput of the ICAP32 DMA which equal to 400 MB/s. We send a word each clock cycle, but due to registers loading and experimental uncertainty, we have an constant overhead of around 80 $\mu$s.

\subsection{DPR power consumption overhead}

The overall resource required to implement the designed system is given in table \ref{tab3}. Moreover, the static and dynamic power consumption of the design mapped on FPGA can be estimated using the Xilinx XPower analyser tool.

\begin{table}[!h]
\renewcommand{\arraystretch}{1.3}
\caption{Total FPGA physical resources usage and power consumption estimation under two configurations}
\label{tab3}
\centering
\begin{tabular}{|c|c|c|}
\hline
\bfseries Resources  &   &  \\
\hline
($sel\_f_1,sel\_f_2$) & (0,1) & (1,1) \\
\hline
Slices  & 2364/7200 & 2427/7200\\
\hline
Registers & 3843/28800 & 4179/28800\\
\hline
LUTS & 4850/28800 & 4998/28800 \\
\hline
\hline
St. power & 450 mW & 450 mW \\
\hline
Dy. power & 256 mW & 270 mW \\
\hline
Total power & 707 mW & 720 mW \\
\hline
\end{tabular}
\end{table}

\begin{figure}[!ht]
    \centering
    \includegraphics[width=0.8\columnwidth]{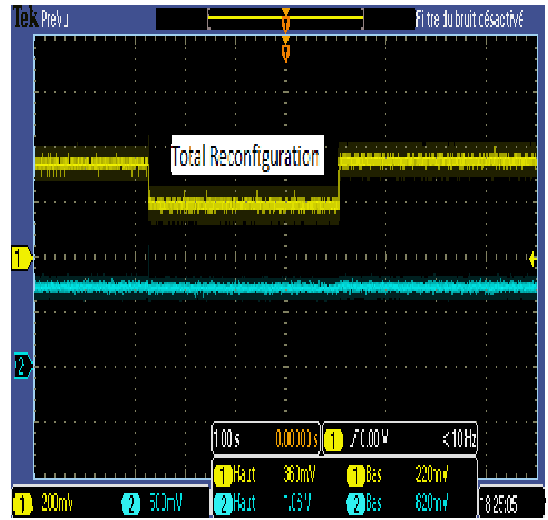}~   
    \caption{FPGA core power consumption during total reconfiguration}
    \label{total}
\end{figure}

\begin{figure}[!ht]
    \centering
    \includegraphics[width=0.8\columnwidth]{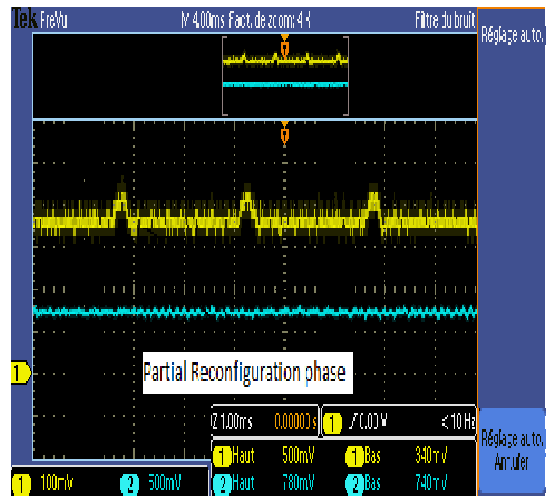}~   
    \caption{FPGA core power consumption during partial reconfiguration}
    \label{partial}
\end{figure}

Fig. \ref{total} and Fig. \ref{partial} show the power consumption of the FPGA core during total reconfiguration and partial reconfiguration processes, respectively. The power consumption overhead is given by VCCINT voltage value. During total reconfiguration, the power consumption of the FPGA core is stable around 220 mW, otherwise it is around 360 mW. During the dynamic partial reconfiguration phase of the APT-VDF filter, the FPGA power consumption overhead overhead doesn't exceed 160 mW (500 mW - 340 mW) during 324 $\mu$s.  

\section{Conclusions} \label{sec:conclusions}
Based on experiment measurement, this paper presents the FPGA core power consumption overhead during the dynamic partial reconfiguration. Three main conclusions can be drawn from our analysis. First, the FPGA dynamic partial reconfiguration is an efficient design technique which permits to change a part of the devices while the rest of an FPGA is still running. Moreover, this technique can reduce dynamic and static power consumption in comparison with the well-known parametrization design approach. Second, the reconfiguration time is very small (around 324 $\mu$s for a partial bitstream size equal to 95Kbytes). Third, as we know, the DPR will introduce power consumption overhead. This work shows that this overhead is small. 

\section{Acknowledgment}
The present work was carried out within the framework of Celtic-Plus SHARING project (number C2012/1-8).
\bibliographystyle{IEEEtran}

\end{document}